\documentclass[12pt]{article}
\usepackage[utf8]{inputenc}
\usepackage{times,placeins}
\usepackage{amsmath,amsthm,amssymb,amsfonts,nccmath,ulem}
\usepackage{graphicx,color,soul,sidecap,caption}
\usepackage[margin=1in,footskip=0.25in]{geometry}

\title{Rapid genetic screening with high quality factor metasurfaces}

\author
{Jack Hu$^{1\ast}$, Fareeha Safir$^{2}$, Kai Chang$^{3}$, Sahil Dagli$^{1}$,\\
 Halleh B. Balch$^{1}$, John M. Abendroth$^{4}$, Jefferson Dixon$^{2}$,\\
Parivash Moradifar$^{1}$, Varun Dolia$^{1}$, Malaya K. Sahoo$^{5}$,\\
Benjamin A. Pinsky$^{5,6}$, Stefanie S. Jeffrey$^{7}$, Mark Lawrence$^{8\ast}$, Jennifer A. Dionne$^{1\ast}$ \\
\\
\normalsize{$^{1}$Department of Materials Science and Engineering, Stanford University}\\
\normalsize{496 Lomita Mall, Stanford, CA 94305, USA}\\
\normalsize{$^{2}$Department of Mechanical Engineering, Stanford University}\\
\normalsize{440 Escondido Mall, Stanford, CA 94305, USA}\\
\normalsize{$^{3}$Department of Electrical Engineering, Stanford University}\\
\normalsize{350 Jane Stanford Way, Stanford, CA 94305, USA}\\
\normalsize{$^{4}$Laboratory for Solid State Physics, ETH Z\"urich}\\
\normalsize{CH-8093 Z\"urich, Switzerland}\\
\normalsize{$^{5}$Department of Pathology, Stanford University School of Medicine}\\
\normalsize{300 Pasteur Drive, Stanford, CA 94305, USA}\\
\normalsize{$^{6}$Department of Medicine, Division of Infectious Diseases and Geographic Medicine,} \\
\normalsize{Stanford University School of Medicine}\\
\normalsize{300 Pasteur Drive, Stanford, CA 94305, USA}\\
\normalsize{$^{7}$Department of Surgery, Stanford University School of Medicine}\\
\normalsize{1201 Welch Road, Stanford, CA 94305, USA}\\
\normalsize{$^{8}$Department of Electrical \& Systems Engineering, Washington University in St. Louis}\\
\normalsize{1 Brookings Drive, St. Louis, MO 63130, USA}\\
\normalsize{$^\ast$To whom correspondence should be addressed; E-mail:}\\
\normalsize{hujack@stanford.edu, markl@wustl.edu,  jdionne@stanford.edu}
}
\date{}

\begin{document}
\maketitle{}
\clearpage
%\linenumbers
\section*{Abstract}
Genetic analysis methods are foundational to advancing personalized and preventative medicine, accelerating disease diagnostics, and monitoring the health of organisms and ecosystems. Current nucleic acid technologies such as polymerase chain reaction (PCR), next-generation sequencing (NGS), and DNA microarrays rely on fluorescence and absorbance, necessitating sample amplification or replication and leading to increased processing time and cost. Here, we introduce a label-free genetic screening platform based on high quality (high-Q) factor silicon nanoantennas functionalized with monolayers of nucleic acid fragments. Each nanoantenna exhibits substantial electromagnetic field enhancements with sufficiently localized fields to ensure isolation from neighboring resonators, enabling dense biosensor integration. We quantitatively detect complementary target sequences using DNA hybridization simultaneously for arrays of sensing elements patterned at densities of 160,000 pixels per cm$^2$. In physiological buffer, our nanoantennas exhibit average resonant quality factors of 2,200, allowing detection of two gene fragments, SARS-CoV-2 envelope (E) and open reading frame 1b (ORF1b), down to femtomolar concentrations. We also demonstrate  high specificity sensing in clinical nasopharyngeal eluates within 5 minutes of sample introduction. Combined with advances in  biomarker isolation from complex samples (e.g., mucus, blood, wastewater), our work provides a foundation for rapid, compact, amplification-free and high throughput multiplexed genetic screening assays spanning medical diagnostics to environmental monitoring. 

\section*{Main}
Genetic screening methods have enabled significant advances in the prediction, detection, treatment, and monitoring of organism and ecosystem health. For example, respiratory panels identify pathogen nucleic acids  indicative of infectious diseases like influenza and Coronavirus disease 2019 (COVID-19)\cite{kim2020,mahony2008}; tissue and liquid biopsies detect cancerous genetic mutations and likelihood of recurrence, and are used to guide treatment\cite{ignatiadis2021,heitzer2018}; and emerging environmental DNA sensors monitor the health of oceans, freshwater, livestock, soil and air\cite{Zhang2021-al,Barnes2016-ry}.  Current genetic screening methods include polymerase chain reaction (PCR), next-generation sequencing (NGS), Sanger sequencing, and DNA microarrays. Each utilizes oligonucleotide amplification followed by optical tagging to sensitively detect target sequences. Despite their tremendous utility in laboratory settings, translation of these screening methods to clinical and point-of-care applications is ultimately limited by their reliance on “traditional” optical signal transduction (absorption and fluorescence). Even with the best optical tags, sensitive and specific readouts are generally only achieved with time consuming thermal cycling and/or costly reagents for nucleic acid amplification.

Nanotechnology-based biosensors have promised new platforms for rapid and scalable bio-molecule detection without requiring biochemical amplification or target labeling. Miniaturized electronic and optical devices offer increased sensitivity due to their nanoscale control of electric and magnetic fields, as well as the potential for scalable multiplexing, owing to their compatibility with complementary metal-oxide-semiconductor (CMOS) fabrication processes. For example,  field-effect transistor (FET) biosensors measure surface potential changes due to molecular binding events \cite{rothberg2011,Nakatsuka2018-iv,seo2020}, while molecular tunnel junction sensors measure changes in tunneling current\cite{reed1997,sorgenfrei2011,porath2000}. These sensors achieve ultra-high sensitivities with femtomolar detection limits, but reliably measuring samples in physiologically-relevant ionic media remains a challenge. 

Complementing electronic sensors, photonic sensors promise high parallelization with more robust read-outs in realistic samples. Rather than amplifying or replicating the biomarker, photonic devices strongly confine and amplify the electromagnetic fields; when decorated with molecular probes, target analyte binding alters the optical signal due to subtle changes in the polarizability or refractive index of the resonator environment. The resonator’s sensing figure of merit (FOM) is generally defined as sensitivity (resonant wavelength shift per refractive index unit (RIU) change) divided by the full width at half maximum (FWHM) of the mode. Plasmonic sensors are among the most common affinity-based biosensors.\cite{Anker2008-qf,Joshi2014-uz,Qiu2020-gv,Im2014-ww,hao2008} These metallic structures exhibit small mode volumes and dipole-like scattering, but generally only achieve FOM values of ca. 1-10 RIU$^{-1}$, due to low quality factors (Q) that are limited by the metals’ intrinsic absorption (Q$\sim$10). Furthermore, due to absorption, these metallic-based structures can lead to sample heating that can degrade biomolecules. 

 More recently, metasurface based sensors have been designed with Q factors of 10’s-100’s, with similar improvements in the FOM\cite{Jahani2021-qb,Bontempi2017-ut,Yavas2019-cq,Tittl2018-xz,Conteduca2021-hg,Yang2014-ta,Wu2014-yn,Wang2021-zr,offermans2011,alam2021,deng2020,liu2020}. Unlike high-Q whispering gallery mode resonators\cite{zhu2010,rodriguez2015,vollmer2008} and photonic crystal microcavity devices\cite{skivesen2007,mandal2009}, these metasurfaces can be illuminated from free space and far field scattering can be readily controlled, an advantage to the scalability and integration of sensors in imaging based devices.\cite{barton2020} However, these systems typically rely on improving Q-factors using delocalized resonant modes formed from extended two-dimensional arrays; the resultant large modal volumes reduce responses to binding of small numbers of target molecules. Additionally, the larger footprint of these arrays ($>$100 by 100 um$^2$) limits the dense incorporation of sensing elements for multiplexed analyte detection and data driven analyses.

In this work, we report a new genetic analysis platform based on our lab's development of high quality factor metasurfaces\cite{Lawrence2020-ew}. These metasurfaces consist of subwavelength nanoantennas that strongly confine light in the near field while affording precise control over the far-field scattering. We design resonators that exhibit high average Q’s of 2,200 in buffered biological media, with strong field penetration into the surrounding environment for sensitive biomarker detection. Due to the spatial localization of the high-Q resonances, individual sensing pixels can be patterned at densities exceeding 160,000 features per cm$^2$, promising detection parallelizability across a multitude of biomarkers. We functionalize our resonators with self-assembled monolayers of DNA probes complementary to the SARS-CoV-2 E and ORF1b gene sequences.  Our sensors rapidly and sensitively detect 22-mer gene fragments within 5 minutes of sample introduction from micromolar to femtomolar concentrations. We further demonstrate high-specificity molecular screening in clinical nasopharyngeal eluates, validating the clinical applicability of our sensor platform for rapid, sensitive, and specific detection of target genes.

\section*{Individually addressable high-Q resonator sensing platform}
Figure 1a illustrates our sensor design, which consists of closely-spaced silicon nanoblocks illuminated with  near-infrared light. Each set of blocks constitutes a one-dimensional  guided-mode resonant (GMR) nanoantenna; the periodic modulation of block widths, characterized by $\Delta$w, allows for finite, but suppressed dipolar radiation and free space coupling to otherwise bound waveguide modes (Supplementary Note 1 and  Supplementary Fig. 3 )\cite{Lawrence2020-ew,Wang1993-iz,Fan2002-en,Overvig2018-do}. The resulting long resonant lifetime translates to strong electric near-field enhancements (Fig. 1b). Notably, electric fields at the surface of Si blocks are enhanced by 80x. Due to the gaps between discrete silicon blocks within the resonator, 29$\%$ of the electric field energy is exposed to the surrounding medium compared with 8$\%$ in a continuous or partially notched waveguide (Supplementary Note 2 and Supplementary Fig. 4). This field concentration in the gaps leads to greater sensitivity to surface-bound analytes. Additionally, these silicon resonators exhibit sharp scattering responses in the far-field. As seen in Figure 1c, calculated reflection spectra Q-factors exceed 5,000 for $\Delta$w=50 nm, and can be further increased with decreased $\Delta$d (\textit{vide infra}).  The high-Q resonances are designed around near infrared wavelengths of 1500-1550 nm to mitigate intrinsic optical absorption due to crystalline silicon and to leverage telecom C-band infrastructure.

We fabricate silicon resonators atop a sapphire substrate (Fig. 1d) (see Methods). Utilizing a near-infrared supercontinuum laser and spectrometer equipped reflection microscope (Supplementary Fig. 1), we illuminate the metasurfaces at normal incidence and simultaneously measure the transmitted spectra from multiple resonators (Fig. 1e). By modulating the block lengths in adjacent nanostructures by $\pm$ 5 nm, we intentionally vary the spectral position of the resonant mode, highlighting that each waveguide structure can be individually addressed and tuned as a distinct resonator (Fig. 1e $\&$ 1f). Our high-Q resonances do not rely on inter-chain coupling or an extended 2-D array effect. This spatial localization of the optical modes makes our platform ideally suited for the integration of densely distributed and multiplexed sensor arrays. 

\section*{Guided-mode resonant metasurface characterization}
Our metasurfaces are sealed in a 3-D printed fluid cell (Fig. 2a) and characterized in phosphate-buffered saline (PBS) solution (1x concentration) to represent physiological conditions for bio-molecule detection. In Fig. 2b, we vary the perturbation $\Delta$w along the block chain from $\Delta$w = 100 nm to $\Delta$w = 30 nm and observe a decrease in the resonant linewidth for at least 25 individual resonators at each condition (Fig. 2b $\&$ 2c). Importantly, in our high-Q metasurface design, the coupling strength between free space radiation and the GMR is dictated by the degree of asymmetry along the waveguide. As silicon is lossless in the near infrared, radiative loss dominates the GMR resonant lifetime and Q factor. Thus, by shrinking $\Delta$w we observe scattering responses with average Q factors of 800 (at $\Delta$w = 100 nm) increasing to 2,200 at $\Delta$w= 30 nm and even observing Q’s above 3,000 for individual resonators (Fig. 2c). 

The observed Q factors represent a two to three order of magnitude increase compared to reported plasmonic biosensors, and a significant ($>$5-10x) increase compared to other non-local metasurface biosensors\cite{Conteduca2021-hg,Yang2014-ta,Wang2021-zr,Yesilkoy2019-wz,Triggs2017-su}, yielding a FOM of $\sim$400 (Supplementary Fig. 7). Our experimental Q factors are slightly lower than numerically predicted likely limited due to scattering losses caused by fabrication imperfections. We also note that water has non-negligible absorption in the 1,500 nm wavelength range that may limit our attainable experimental Q factors (Supplementary Note 3 and Supplementary Fig. 5). Designing future resonators in an optical transparency window of biological media (such as 1,300 nm) and optimizing fabrication processes may further improve performance, with Q factors in the millions potentially attainable; such structures could offer the single particle sensitivity of high-Q microcavities,\cite{zhu2010,vollmer2008} but with the ease of integration and compact form-factor afforded by free-space coupling.

Due to the localization of the mode along each individual chain, resonators can be spaced laterally at least as close as 3 $\mu$m without affecting the GMR (Fig 2d). Based on our fabricated waveguide length of 200 $\mu$m, our devices feature sensor arrays with densities of over 160,000 sensors per cm$^2$. Due to the slow group velocities of the GMR’s, losses due to finite size effects can be suppressed\cite{Overvig2018-do,Mizutani2003-oz}, and 50 $\mu$m waveguides can be fabricated with comparable Q (Supplementary  Fig. 6), yielding feature densities over 600,000 sensors per cm$^2$. These large sensor densities offer an avenue for robust statistical analysis in diagnostic studies as well as a platform for multiplexed detection of many distinct biomarkers in parallel.

\section*{Self-assembled monolayer functionalization and sensing}
To utilize our sensor arrays for nucleic acid detection, we modified the silicon surface with DNA monolayers, where complementary nucleic acid sequences serve as capture molecules for a specified target genetic sequence. Self-assembled monolayers (SAMs) are deposited in a three-step process to covalently link 22 base pair single-stranded  DNA (ssDNA) probes over the entire metasurface chip surface. The silicon surface is first functionalized with an amine-terminated silane (11-aminoundecyltriethoxysilane, AUTES), and then cross-linked via a heterobifunctional molecule (3-maleimidobenzoic acid N-hydroxysuccinimide ester, MBS) to thiolated ssDNA probes (Methods and Supplementary Fig. 2). In this study, we considered nucleic acid fragment targets of the envelope (E) and open reading frame 1b (ORF1b) genes of the SARS-CoV-2 virus (GenBank accession: MT123293.2 positions 26326→ 26347 and 18843→ 18866, respectively, also see Supplementary Table 1)\cite{Corman2020-ec,Chu2020-jl}(Fig. 3a). As a proof of principle, we use synthetic DNA targets, but note that viral RNA will analogously hybridize to complementary DNA probes\cite{Rauzan2013-yn,Cheung2020-xk}. 

In Fig. 3b, measured spectra show clear shifts to the resonant wavelength as consecutive molecular monolayers of AUTES, MBS, and the probe DNA are grafted to the resonator surface. In numerical simulations, monolayers were modeled as thin dielectric shells surrounding the silicon blocks and simulated responses show close agreement with the experimental resonance shifts (Fig. 3c)(Supplementary Note 6 and Supplementary Fig. 8). Upon adding a solution of target SARS-CoV-2 gene fragments, a clear resonant shift is observed (Fig 3d). Data was collected from N=75 individual resonators. The high density of sensing elements on our chips can enable significant increases in measurement throughput compared to typical photonic sensors where signals are averaged over larger 2-D arrays. The deviation between experimental and simulated wavelength shifts for the AUTES and MBS layers is likely due to the tendency for aminosilane molecules to form multilayer structures; differences in the attachment of DNA probes and subsequent target hybridization are likely due to a strong influence of steric hindrance and electrostatic repulsion effects on the packing density and hybridization efficiency of the DNA strands\cite{Seitz2011-xu,Zhu2012-uy,SantaLucia1998-ov,Takashima2015-xy}.

\section*{Rapid and specific gene fragment detection}
Pairing our resonators with specific probe DNA sequences offers specificity in target gene detection. We modify our surface chemical functionalization process with an antifouling matrix to reduce non-specific binding signals. A 1:1 mixture of thiolated ssDNA probes and thiolated polyethylene glycol (PEG) chains is immobilized on the silicon nanostructures, where PEG has been shown to mitigate biofouling\cite{guardado2021,sharma2004}. To confirm specificity, we modify target DNA strands with ATTO590 fluorescent labels and incubate sensors functionalized with probes that are only complementary to the nCoV.E sequence. Fluorescence imaging of sensors exposed to 1 $\mu$M solutions of target nCoV.E and HKU.ORF1 show significant binding only for the complementary E gene target and minimal signal for the non-complementary ORF1 strands (Supplementary Fig. 10). This target specificity is also measured in the resonator scattering spectra, where resonance wavelength shifts are significant for complementary target-probe conditions and suppressed for non-specific binding (Fig. 4a). 

Our sensors exhibit concentration dependent responses from 1 $\mu$M to 1 aM (Fig. 4b). Measurements are taken for N=20 individual resonators at each target and concentration condition. As seen in Fig. 4b, experimental resonant shifts vary from 0.2 nm at 1 $\mu$M to 0.01 nm at 1 aM. The concentration curve of nCoV.E targets is fit to the Langmuir adsorption model, which is used to describe target binding coverage in affinity-based assays (Supplementary Note 8). We estimate the limit of detection (LOD) to be $\sim$8 fM, based on the IUPAC (International Union of Pure and Applied Chemistry) definition, which is the sum of the mean blank measurements and 3X the standard deviation of the blank measurements (LOD = $\mu$+3$\sigma$). The LOD observed from our devices represents a significant improvement compared to previous nanophotonic nucleic acid sensor studies\cite{Qiu2020-gv,Joshi2014-uz,hu2014,rong2008,qavi2010}. Furthermore, this LOD corresponds to approximately 4,000 copies/$\mu$L, which is on the order of clinically measured viral loads (10$^3$-10$^5$ copies/$\mu$L) in infected patients\cite{pan2020,savela2022}. With a detection limit in the low femtomolar regime, our sensor is promising for amplification-free and label-free viral diagnostics. We note that the target nucleic acids used in this study are only 22 base pairs in length; optimization of sensors for longer gene fragment targets could further reduce the LOD as larger molecules produce a stronger perturbation to the local refractive index. Further, the concentration dependent range of our device can potentially be tuned to different values of analyte concentration through modification of surface probe densities \cite{Nakatsuka2018-iv}.

Efficient free-space scattering from our metasurface resonators enables real-time measurements of target binding. In Fig. 5a we show the time dependent measurement of 10 resonators with spectra acquired at 5 second intervals for concentrations of 1 $\mu$M, 1nM, 1pM, and 1fM. For all concentrations, we observe resonant wavelength shifts within seconds of target injection. For 1$\mu$M, 1nM, and 1pM binding curves we see signal saturation within 10 minutes as the DNA hybridization process reaches dynamic equilibrium. The signal response shows excellent agreement with the Langmuir adsorption model (dashed line Fig. 5a) where the observed hybridization rate constants of 10$^{-3}$ to 10$^{-2}$ s$^{-1}$ are comparable to other hybridization capture assays\cite{Xu2017-mt,Gao2006-if,vanjur2020}. These fast binding kinetics highlight a key advantage of chip-based approaches over conventional detection techniques that require time-intensive ($\sim$2-8 hours) molecular amplification cycles.

We further validate the performance of our metasurface sensor with clinical nasopharyngeal swabs. As a proof-of-principle demonstration, we utilize nasopharyngeal eluates from nasopharyngeal swab specimens submitted to the Stanford Clinical Virology Laboratory for SARS-CoV-2 reverse transcription polymerase chain reaction (RT-PCR) testing. The specimens were tested by a laboratory developed FDA-EUA approved RT-PCR assay targeting the E-gene. Total nucleic acid was extracted from 400 $\mu$L of the swab resuspended in viral transport medium or PBS using QIAsymphony DSP Virus/Pathogen Midi Kit in QIAsymphony automated platform (Qiagen, Germantown, MD) and eluted in 60 $\mu$L buffer AVE containing $\sim$60 ng/$\mu$L of carrier RNA. This study used SARS-CoV-2 negative eluates which represent sample media containing human and other nucleic acids including a high concentration milieu of poly-A or random nucleic acids of various lengths. In Figure 5b, we flow the nasopharyngeal eluates over our sensor and show that there is minimal sensor response (N=10 resonators) in negative samples, despite the background matrix of non-specific biomolecules present in clinical samples. Subsequent rinsing with pure PBS buffer returns the resonance wavelengths to baseline values. We then inject a nasopharyngeal sample solution spiked with 100 nM of complementary nCoV.E target molecules. A clear resonance wavelength shift of 0.25 nm occurs within 5 minutes; the complementary target signal remains stable as the sensor is rinsed with clean buffer solution.  Therefore, our sensor is able to discriminate between specific target molecules and non-specific binding signals even in complex clinical samples, demonstrating the robustness of our self-assembled monolayer functionalized nanostructures.

\section*{Conclusions}
Our nanophotonic device offers a new platform for high throughput molecular analysis. We have demonstrated free space illuminated resonators with high-Q resonances in physiological media (2,200+) that can be patterned, tuned, and measured at densities exceeding 160,000 pixels per cm$^2$. Even larger Q’s and greater feature densities can be obtained in our platform with improved fabrication processes to reduce scattering losses from structural inhomogeneities, reduced absorption losses from biological media, and inclusion of photonic mirror elements to suppress light leakage as resonator chains are truncated below 50 $\mu$m. Interfaced with DNA probes, our metasurface design enables rapid, label-free, and highly digitized genetic screening that can bridge many of the challenges faced by conventional genetic analysis techniques. This increased digitization of target gene binding may also be integrated with machine learning based analysis for further improved accuracy or to allow for discrimination of small signals due to genetic variants and point mutations\cite{Cui2020-on}. Paired with bioprinting procedures where different gene sequence probes are spotted across distinct sensing pixels, our high-Q metasurface chips can provide the foundation for rapid, label-free, and massively multiplexed photonic DNA microarrays. Furthermore, our nanophotonic chips are amenable to intensity imaging and/or hyperspectral imaging techniques that provide signal binding information without the need for a spectrometer\cite{Tittl2018-xz,Yesilkoy2019-wz}, further reducing complexity and costs towards point of care genetic screening. Our platform promises unique possibilities for widely scaled and frequently administered genetic screening for the future of precision medicine, sustainable agriculture, and environmental monitoring.

\begin{figure}[ht!]
\includegraphics[width=16 cm]{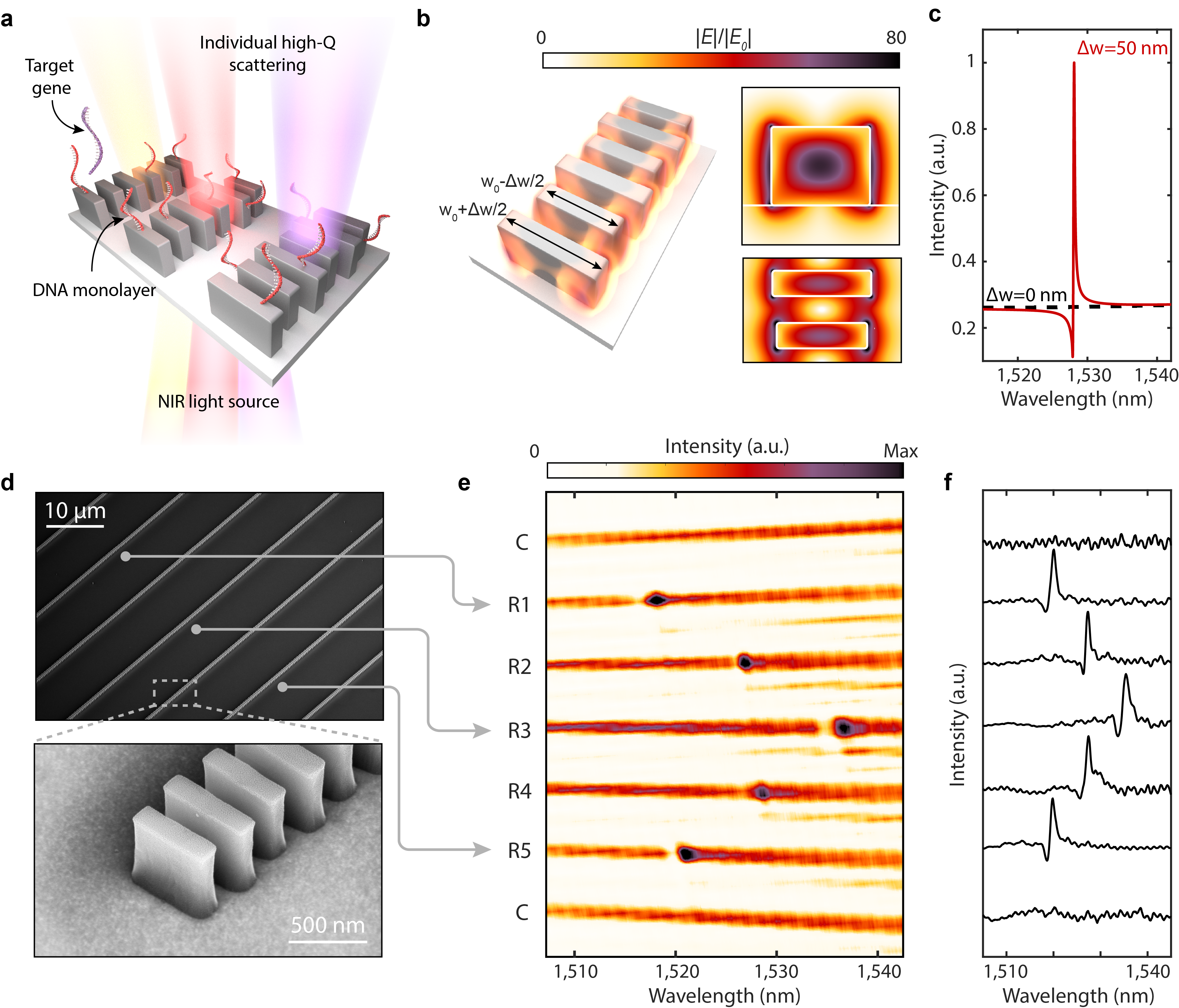}
\caption{\label{fig1}\textbf{Design of high-Q sensors.} \textbf{a}, Metasurface arrays of high-Q guided mode resonators consisting of perturbed chains of silicon blocks interfaced with DNA probes for targeted gene detection. Geometrical parameters of the resonators are height (h) = 500 nm, $w_0$ = 600 nm, thickness (t) = 160 nm, block spacing ($a_y$ = 330 nm), inter-chain spacing ($a_x$ = 10 $\mu$m), and $\Delta$w varied between 30-100 nm.  \textbf{b}, Simulated electric near-field enhancements for a  resonator with $\Delta$w = 50 nm. \textbf{c}, Simulated cross-polarized transmission response of metasurface illuminated with normally incident linearly polarized plane waves. Responses normalized to intensity maximum of perturbed resonator. \textbf{d}, SEM micrographs of metasurface device composed of multiple individually monitored and tuned resonators. \textbf{e}, Spectral image from array with 7 resonators where C denotes nanostructures with no perturbation $\Delta$w = 0 nm and R1-R5 having perturbation $\Delta$w = 50nm. Resonance positions are modulated by adjusting block length where $w_0$ = 595 nm for R1 $\&$ R5, $w_0$ = 600 nm for R2 $\&$ R4, and $w_0$ = 605 nm for R3 to form the observed chevron pattern. \textbf{f}, Row averaged transmitted intensities corresponding to \textbf{e}.}
\end{figure}

\begin{figure}[ht!]
\centering
\includegraphics[width=12cm]{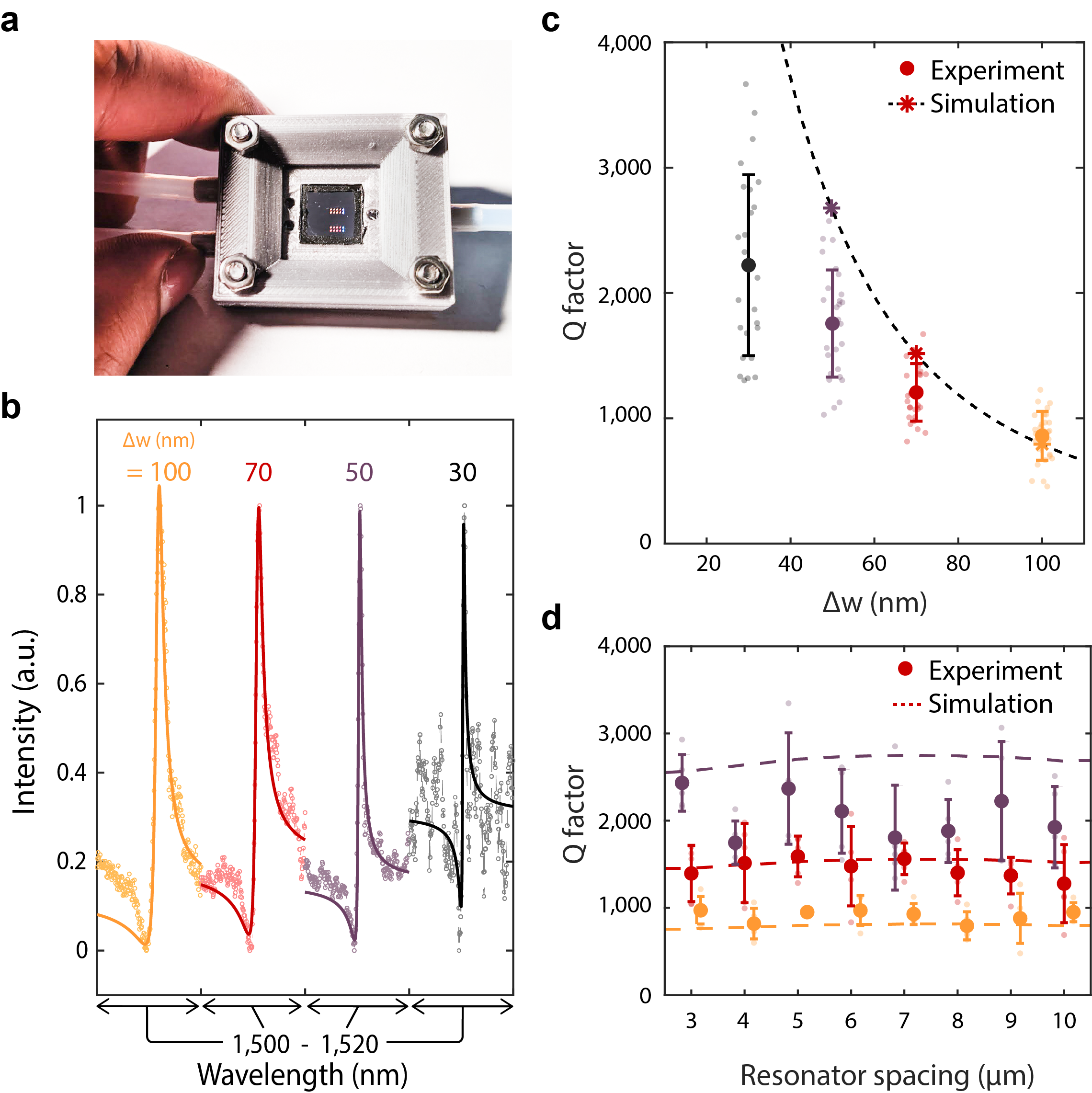}
\caption{\label{fig2}\textbf{Fluid cell characterization of metasurfaces}. \textbf{a}, Photo of metasurface chip enclosed in fluid cell. \textbf{b}, Representative spectra from resonators with varying $\Delta$w. Solid lines represent fits to a Lorentzian oscillator. \textbf{c}, Quality factor of resonances with different $\Delta$w. Bold markers and error bars are the mean and standard deviation for N=30 resonators at each condition. Stars represent simulated values and the dashed line is a fit to predicted values from coupled mode theory (Supplementary Note 3). \textbf{d}, Quality factor as a function resonator spacing where mean and standard deviation are for N=5 resonators at each condition.}
\end{figure}

\begin{figure}[ht!]
\includegraphics[width=16 cm]{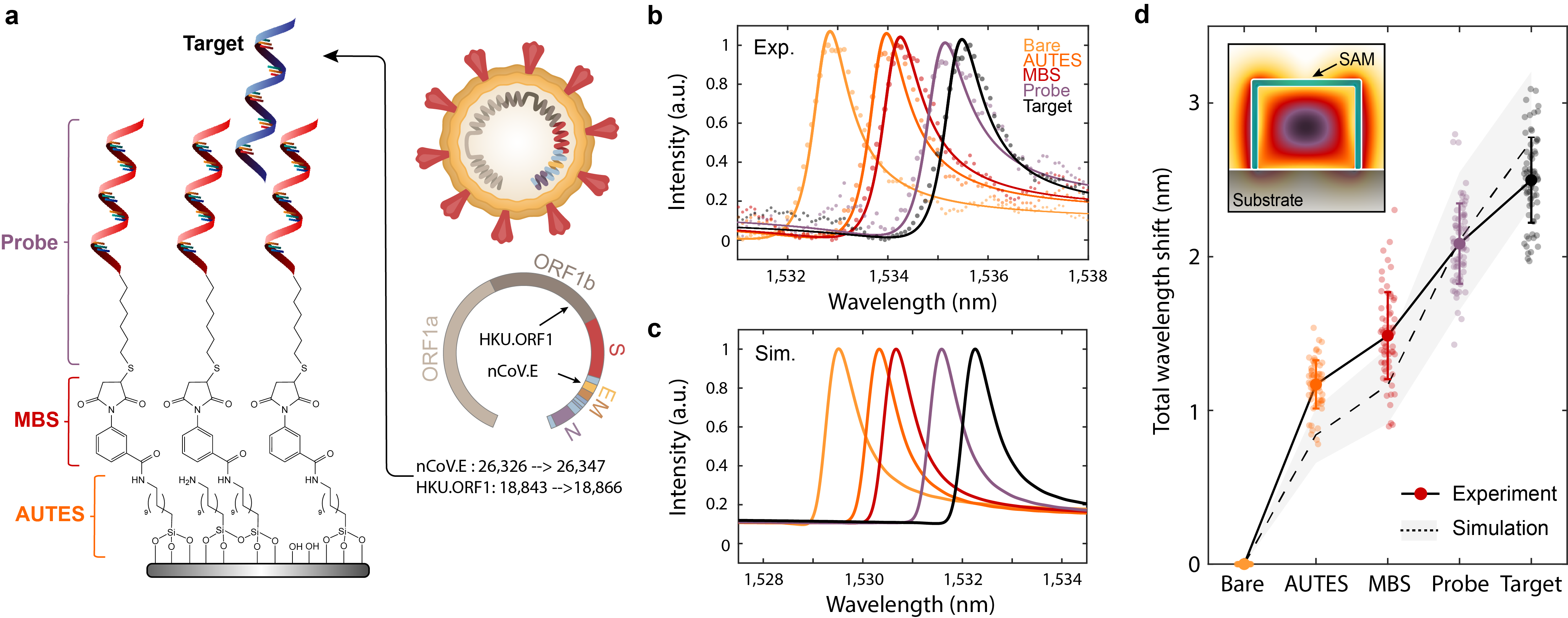}
\caption{\label{fig3}\textbf{DNA monolayer functionalization and resonant wavelength shift measurement}. \textbf{a}, Schematic of chemical components utilized in immobilizing DNA self-assembled monolayers (SAM) onto the silicon nanostructures. Target DNA fragments for this study are portions of the E and ORF1b genes from the SARS-CoV-2 virus. \textbf{b}, Experimentally measured and \textbf{c}, simulated resonance wavelength shift responses with the addition of each molecular layer in the SAM, including complementary nCoV.E target binding. Markers in \textbf{b}, correspond to measured data points while solid lines show fits to a Lorentzian oscillator. The difference in absolute wavelength values between experimental and simulated spectra can be attributed to slight dimension variations in the fabricated structures. \textbf{d}, Total resonant wavelength shift during SAM functionalization and DNA sensing as referenced from initial measurements on bare silicon structures. Markers represent individual measurements from N=75 independent resonator devices and bolded markers and error bars are the mean and standard deviation of the measurements.}
\end{figure}

\begin{figure}[ht!]
\centering
\includegraphics[width=12 cm]{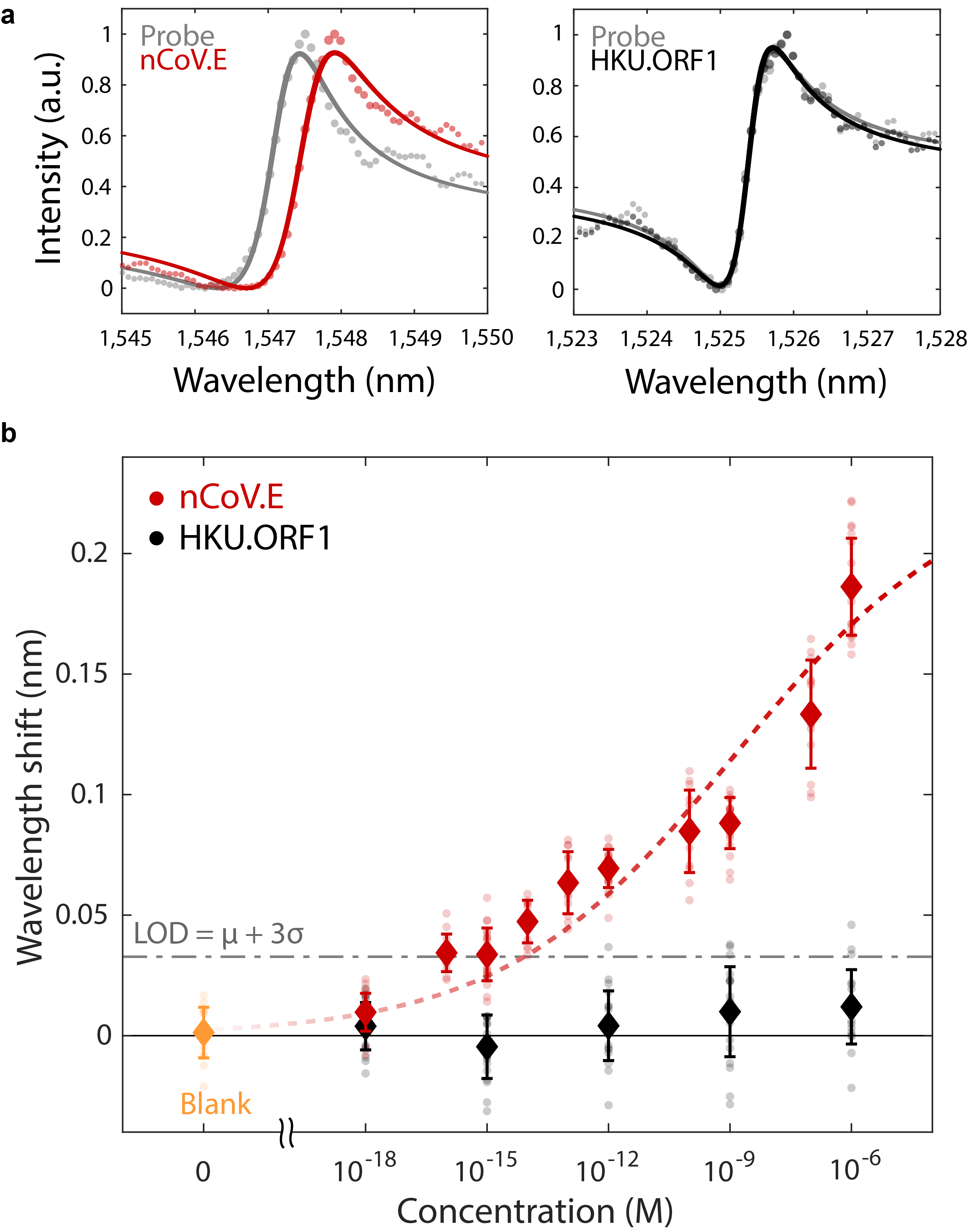}
\end{figure}
\begin{figure}[tp!]
\caption{\label{fig4}\textbf{Biosensing demonstration with SARS-CoV-2 gene fragment targets}. \textbf{a}, Measured spectra from individual resonators indicate significant wavelength shifts of $\sim$0.2 nm with complementary DNA binding and minimal signal changes when introduced to non-complementary sequences. \textbf{b}, Concentration dependent binding responses for both nCoV.E and HKU.ORF1 targets incubated on metasurface devices functionalized with only nCoV.E complementary probes. Error bars indicate standard deviations of measurements from N= 20 measurements from distinct resonators for each target and concentration condition. The limit of detection is estimated based on the mean + 3 standard deviations of the blank measurements. Dashed lines show fits to the Hill equation (Supplementary Note 8).}
\end{figure}

\begin{figure}[ht!]
\centering
\includegraphics[width=9 cm]{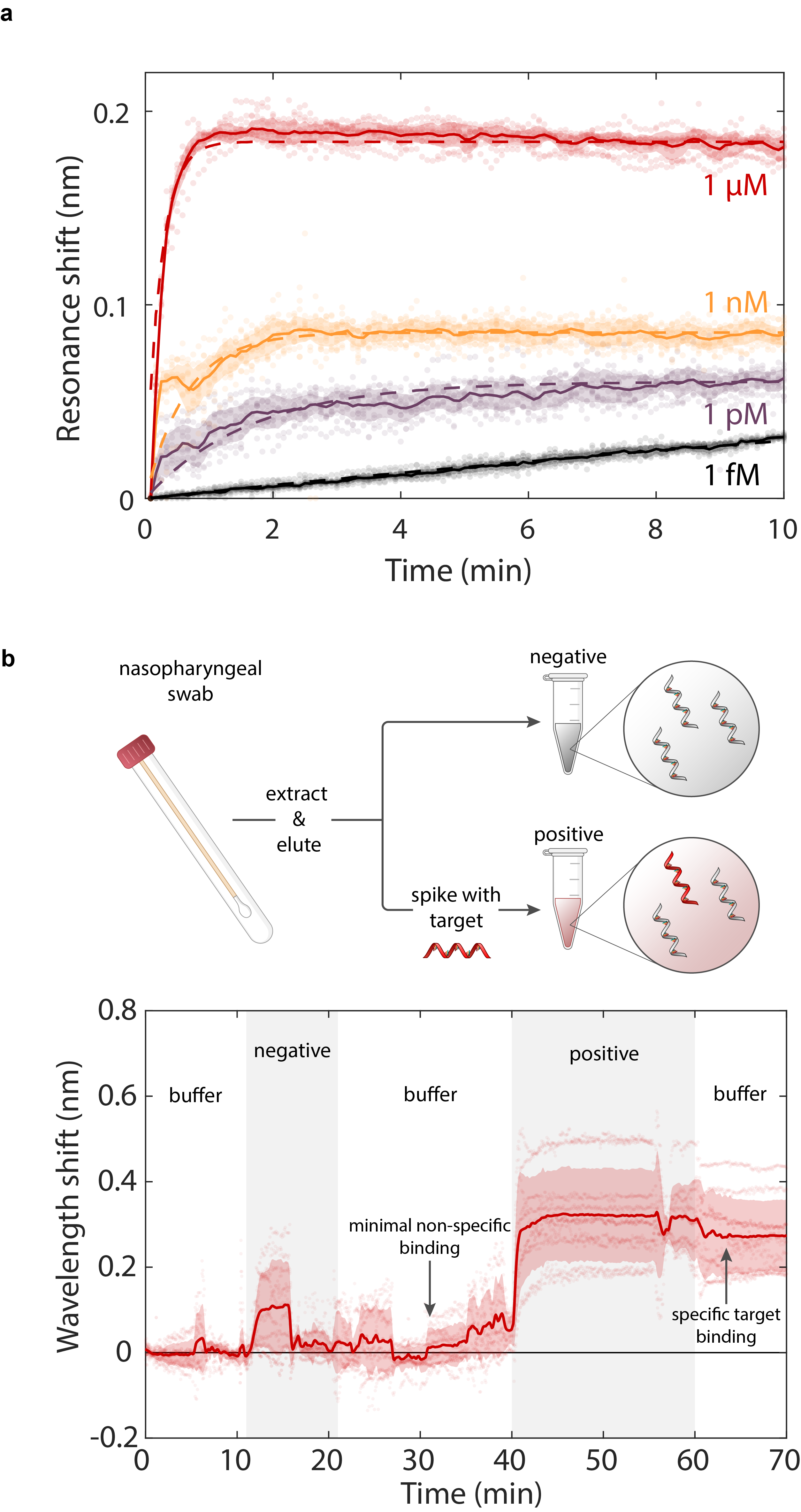}
\end{figure}
\begin{figure}[tp!]
\caption{\label{fig5}\textbf{Kinetic binding response and measurement in clinical nasopharyngeal samples}. \textbf{a}, Time-dependent binding responses from 10 distinct resonators exposed to 1 fM, 1pM, 1nM, and 1$\mu$M concentrations of nCoV.E target molecules. \textbf{b}, Demonstration of gene fragment detection in clinical nasopharyngeal eluates. Negative samples contain random and scrambled genetic material from nasopharyngeal swabs that have been confirmed negative for SARS-CoV-2 via RT-PCR. Target nCoV.E molecules are spiked into the negative nasopharyngeal eluates at a concentration of 100 nM for the positive sample.
}
\end{figure}

\FloatBarrier
\pagebreak
\bibliographystyle{naturemag}
\bibliography{references}

\section*{Methods}
\subsection*{Computational design}
Electromagnetic simulations were performed with the Lumerical FDTD Solver. Metasurfaces were simulated with periodic boundary conditions in the x and y directions and perfectly matched layer (PML) boundary conditions in the z direction. Structures were excited with a plane wave polarized at 45$^\circ$ and injected from the negative z direction through a sapphire substrate. Transmission spectra were computed using a power monitor placed in the far field of the metasurface in the +z direction. Cross polarized transmission intensity was calculated as Power($-45^\circ$)/(Power($-45^\circ$)+ Power(+45$^\circ$)).

\subsection*{Fabrication}
The metasurfaces were fabricated using standard lithographic procedures. First, 500 nm, single crystal silicon-on-sapphire (MTI Corp.) substrates were cleaned via sonication in acetone and isopropyl alcohol. The substrates were baked at 180 $^\circ$C before spin coating with hydrogen silsesquioxane (HSQ) negative tone resist (XR-1541-06, Corning). The resist was baked for 40 min at 80 $^\circ$C. To reduce charging, a charge dissipation layer (e-spacer, Showa Denko) was spin coated over the HSQ resist and baked again for 5 min at 80 $^\circ$C. The metasurface patterns were defined by a 100 keV electron beam in a JEOL JBX-6300FS EBL system. Patterns were developed for 120 seconds in a 25$\%$ solution of tetramethylammonium hydroxide. Reactive ion etching with Cl2, HBr, and O2 chemistries were utilized to transfer the pattern to the silicon layer (Lam TCP 9400). The HSQ resist was removed using 2$\%$ hydrofluoric acid in water and the samples were then cleaned using a Piranha solution (9:1 H$_2$SO$_4$:H$_2$O$_2$) heated to 120 $^\circ$C. The silicon nanostructures were passivated by heating for 30 min at 800 $^\circ$C in a furnace to grow a $\sim$ 4 nm oxide layer.

\subsection*{Optical characterization}
Resonator spectra were measured in a home-built near-infrared reflection microscope shown in Supplementary Figure 1. Samples were illuminated via a broadband NKT supercontinuum laser with a collimated fiber output. A polarizer P1 was set to create linearly polarized incident illumination at a 45$^\circ$ angle with respect to the metasurface structures. The illuminating beam is focused to the back focal plane of a 5X objective (Mitutoyo Plan Apochromat NIR) with a lens L1 (f=100 mm) to produce a collimated plane wave at the sample. The devices were illuminated through the sapphire substrate. Additionally, all optical measurements in this work were taken with sample chips sealed in a fluid cell and immersed in PBS 1X. The scattered light is  directed through a cross-polarized polarizer P2 at -45$^\circ$ to reduce the substrate Fabry-Perot signal. The scattered light is then focused via a lens L3 (f=75 mm) into a SPR-2300 spectrometer (Princeton Instruments). The broadband signal is diffracted via a diffraction grating (600 g/mm, blase wavelength 600 nm, Princeton Instruments) and focused onto an air-cooled InGaAs detector (NiRvana, Princeton Instruments). Spectral features were analyzed by fitting the data with the function:

\begin{equation*}
    T = \left| a_r + a_i i + \frac{b}{f-f_0+\gamma i}\right|^2
\end{equation*}

where $T$ is the scattered intensity from a superposition between a constant complex background, $a_r+a_i i$, and a Lorentzian oscillator with resonant frequency $f_0$ and full-width at half-maximum of 2$\gamma$. The quality factor is then calculated as $Q = f_0/2\gamma$.

\subsection*{Surface functionalization}
 Self-assembled monolayers of single stranded probe DNA was interfaced to the silicon metasurfaces through a multi-step chemical functionalization process summarized in Supplementary Figure 2. To activate the silicon surface for functionalization, the samples were immersed in a Piranha solution (9:1 H$_2$SO$_4$:H$_2$O$_2$) heated to 120 $^{\circ}$C for 20 min to hydroxylate the surfaces. Next, samples were immersed in a 0.1 mM solution of  11-aminoundecyltriethoxysilane (Gelest Inc.) in ethanol, sealed, and left for overnight for 18-24 hrs. The samples were rinsed in fresh ethanol for 5 min (3X) and then baked for 1 hr at 150 $^{\circ}$C to form a stable silane layer. A hetero-bifunctional cross linking molecule was attached to the silane layer through immersion in a 1mM solution of 3-maleimidobenzoic acid
N-hydroxysuccinimide ester (Millipore Sigma) dissolved in a 1:9 (v/v) mixture of dimethyl sulfoxide and PBS for 1 hr. Samples were then rinsed thoroughly with deionized water and blown dry with N$_2$ gas. Single stranded DNA probes were obtained from Integrated DNA Technologies (Coralville, IA) modified with a disulfide tether on the 5' ends. The as received DNA probes were disperesed in 50 $\mu$L of tris-EDTA buffer, pH 8.0, and mixed with 30 mg of DL-dithiothreitol for at least 1 hr to reduce the disulfide moieties to thiols. The probes were then purified via gravity-flow size exclusion chromatography using illustra NAP-5 columns. The concentration of the eluted DNA solutions were determined using UV absorption signatures (Varian Cary 500 UV-Vis Spectrophotometer). For the functionalization reaction, portion of the stock solution were then diluted to 20 $\mu$M in PBS 1x with added divalent cations of 100 mM MgCl$_2$. For measurements presented in Fig. 4 and Fig. 5, the DNA probes were mixed in a 1:1 ratio (10$\mu$M:10$\mu$M)  with thiolated monomethoxy polyethylene glycol (mPEG) (MW = 350) from Nanocs (Boston, MA). The DNA probe solution was pipetted onto each sample and incubated overnight ($\sim$18-24 hrs) in a dark and humid environment.  Samples were rinsed with PBS 1X and then soaked in a PBS solution with added salt to a concentration of 1M NaCl for 4 hours to remove any loosely bound or physiosorbed oligonucleotides. Samples were then rinsed with PBS 1X and deionized water and dried with N$_2$ gas. Samples corresponding to optical measurements in main text Fig. 3 were measured before and after each functionalization step with additional deionized water rinsing and N$_2$ drying before the next chemical processing step. Samples corresponding to main text Fig. 4 adn 5 were optically characterized only before and after target DNA hybridization. 

\subsection*{DNA hybridization}
For DNA hybridization measurements presented in Fig. 4, a baseline spectroscopic measurement was taken on metasurfaces that had been functionalized with a probe DNA monolayer. Probes with sequences corresponding to the E gene of the SARS-CoV-2 virus were used in all experiments. Following baseline measurements, samples were exposed to a solution containing either E or ORF1b gene fragments. A target DNA solution corresponding to either complementary E gene or non complementary ORF1b gene fragments (Supplementary Table 1) was produced by diluting a 100 µM stock solution to the desired concentration in 1X PBS. Additional divalent cations corresponding to 100 mM MgCl$_2$ were added to the solution to increase hybridization efficiency and speed. The resonant wavelength shift for each concentration was recorded after 20 minutes of incubation.

For dynamic DNA hybridization measurements presented in Fig. 5 of the main text, samples functionalized with DNA probes were placed in a fluid cell and mounted in the optical transmission set up described above. Spectral acquisitions were collected at 5 second intervals, and measurements were conducted under a continuous flow rate of 50 $\mu$L/min. 

For measurements presented in Fig. 5b, nasopharyngeal swabs were collected from clinical patients at the Stanford Clinical Virology Laboratory. Nasopharyngeal swabs were submitted for SARS-CoV-2 RT-PCR testing and were tested by an FDA-EUA approved RT-PCR Assay targeting the E-gene. Total nucleic acid was extracted from 400 $\mu$L of the swab resuspended in viral transport medium or PBS using QIAsymphony DSP Virus/Pathogen Midi Kit in QIAsymphony automated platform (Qiagen, Germantown, MD) and eluted in 60 $\mu$L buffer AVE containing $\sim$60 ng/$\mu$L of carrier RNA. Collected samples were pooled to create the negative control sample. Synthetic E gene fragment targets were spiked into the pooled negative control samples at a concentration of 100 nM to create the positive test samples.

\section*{Acknowledgments}
The authors thank Elissa Klopfer, Dr. Harsha Reddy, Dr. Loza Tadesse, Dr. David Barton, Dr. Lisa Poulikakos, Dr. Chris Siefe, and Baba Ogunlade for insightful discussions. The authors acknowledge funding from a NSF Waterman Award (Grant number 1933624), which supported the salary of J.A.D., the Air Force Office of Scientific Research (Grant FA9550-20-1-0120), which supported the chip design and fabrication and salary of M.L., and the NIH New Innovator Award (1DP2AI152072-01), which supported the surface functionalization and sensor characterization. J.H. and J.D. were also supported by seed funds from the Stanford Center for Innovation in Global Health, an innovation transfer grant from the Stanford TomKat Center, and a translation grant from the Stanford Byers Center for Biodesign via the Spectrum MedTech Pilot Project program. H.B.B. was supported by the NIH New Innovator Award (1DP2AI152072-01) and the NSF OCE-PRF (Grant Number 2205990) Part of this work was performed at the Stanford Nano Shared Facilities and Stanford Nanofabrication Facilities, which are supported by the National Science Foundation and National Nanotechnology Coordinated Infrastructure under awards ECCS-2026822 and ECCS-1542152. 

\section*{Author contributions}
J.H., M.L, J.M.A, and J.A.D. conceived and designed the experiments. J.H. and M.L. conducted the theory and numerical simulations. J.H., S.D., P.M., and J.D. fabricated the nanostructured samples. J.H., F.S., K.C., and J.M.A. performed the surface functionalization of the sensors. J.H., F.S. K.C., H.B.B., and V.D. performed the optical characterization experiments. M.K.S. processed the clinical nasopharyngeal samples. J.A.D. conceived the idea and supervised the project, along with M.L., S.S.J. and B.A.P. on relevant portions of the research. All authors contributed to the preparation of the manuscript. 

\end{document}

\maketitle
\clearpage

\section*{Supplementary Note 1: Dispersion calculations}

Considering an unperturbed silicon waveguide made from a 1-D array of subwavelength silicon blocks, we calculate the waveguide dispersion for the lowest order mode (Supplementary Fig. \ref{figs3}a). The waveguide mode possesses larger momentum than free-space radiation and is ”bound” and does not couple to free-space illumination. Upon introducing periodic perturbations in the length of every other silicon block along the waveguide, our unit cell spacing, a, effectively doubles, folding the first Brillouin zone in half. Now modes that were previously inaccessible to freespace illumination lie above the light line (Supplementary Fig. \ref{figs3}b). This band structure is maintained when the magnitude of the perturbation is changed and only the coupling strength and Q factor are modulated as discussed in the main text.

\section*{Supplementary Note 2: Spatial distribution of electric fields around resonators}

The sensitivity of a resonant mode to minute changes in the local refractive index can be estimated by the fraction of electric field energy residing outside the resonator. We calculate the exposure of the mode utilized in our sensors with the following equation:

\begin{center}
\begin{equation*}
        f_{U_{E}} = \frac{\int_{V_{out}}\epsilon_{out}|E|^2dV_{out}}{\int_{V_{in}}\epsilon_{in}|E|^2dV_{in}}
\end{equation*}
\end{center}

where $\epsilon_{out}$ and $\epsilon_{in}$ are the permittivity of the medium containing the analyte and the permittivity of the resonator and substrate, respectively. $V_{out}$ and $V_{in}$ represent the volumetric regions of the analyte containing medium and the portions inside the resonator or substrate that do not overlap with any bound materials or molecules. Performing this analysis on the sensor design described in the main text as well as guided mode resonant structures previously described in reference \cite{Lawrence2020-ew} composed of notched silicon waveguides, we find that our silicon block chains significantly increase field penetration into the surrounding environment. Field profiles of the two structures are plotted in Supplementary Fig. \ref{figs4} showing similar transverse electric waveguide modes. Due to the subwavelength spacing of the discrete silicon blocks in our sensors, we still excite the localized waveguide modes along the periodic direction that are seen in continuous silicon wire waveguides. However, the grating-like structure exposes regions of the mode to the surroundings while also reducing the effective mode index of the waveguide, leading to further extension of the fields out of the resonator. This design results in the fraction of the mode energy in the surroundings to increase to $f_{U_{E}}=0.29$ compared to only $f_{U_{E}}=0.08$ for notched or continuous waveguide structures.

\section*{Supplementary Note 3: Quality factor scaling and water absorption}

As discussed in the main text, introducing an asymmetry along a silicon waveguide allows for the excitation of previously bound modes. Reduction of the asymmetry, $\Delta$w in the case of our metasurfaces, decreases the coupling strength of the mode to free-space radiation thereby increasing the Q factor. For a material that exhibits no intrinsic absorption losses, such as silicon in the near infrared, the Q factor can be arbitrarily increased as the perturbation strength approaches zero. This dependence of the Q factor on subtle structural deviations have been previously described through temporal coupled-mode theory and perturbation theory\cite{Fan2002-en,Overvig2018-do,koshelev2019}:

\begin{center}
    \begin{equation*}
        Q=\frac{B}{\alpha^2}
    \end{equation*}
\end{center}

where B is a constant that depends on the resonator geometry and $\alpha$ is a unit-less asymmetry parameter represented by $\Delta$w/w$_0$ in our metasurface. This relationship is shown in Supplementary Fig. \ref{figs5}, where theory (solid line) and numerical simulations (stars) indicate diverging Q factors as $\Delta$w is decreased. We also observe that experimentally observed Q factors are lower than predicted values (experimental data from Main text Fig. 2). One significant factor limiting our experimental quality factors is the absorption coefficient of water at telecommunication wavelengths. Since all our optical measurements are performed in aqueous solutions, dissipative losses are expected to decrease our measured Q factors as shown by the dashed line in Supplementary Fig. \ref{figs5}, which represents numerical calculations including water absorption. The effects of absorption losses are particularly strong as $\Delta$w is decreased, as longer resonance lifetimes lead to greater interaction between the resonant mode and the absorptive background medium. Future iterations of our sensor can be designed in the water absorption window around 1300 nm to maximize performance of the resonators. Additionally, fabrication imperfections such as surface roughness or non-uniformity in the metasurface structures will introduce scattering losses and reduce the observed Q factor.

\section*{Supplementary Note 4: Finite size resonators}

While the resonators shown in the main text exhibit high-Q modes in longer 1-D arrays (200 $\mu$m), we show that the resonators can be scaled down significantly while maintaining sharp spectral features. Our metasurface design features low scattering losses out the ends of the waveguides, and hence are relatively robust to resonator finite size effects due to the high index contrast between separated silicon blocks and gaps containing the background medium. In Supplementary Fig. \ref{figs6}a, we show calculated dispersion diagrams for three different resonators consisting of a solid silicon waveguide with increasing depths of notch corrugations. The waveguide has width of 600 nm and from top to bottom, the bands correspond to notches added on both sides of the waveguide with depths of 50, 150, and 300 nm. We observe flattening of the bands as the notch depth is increased until 300 nm, where the waveguide is now separated into distinct silicon blocks. The flatter bands indicate a much smaller group velocity due to strong in-plane Bragg scattering, which reduces the propagation of the mode out the waveguide ends and reduces effects of shrinking the resonator on the Q factor.
We experimentally verify that we can maintain high quality factors while shortening the overall length of each resonator. In Supplementary Fig. \ref{figs6}b,c we show SEM images of multiple resonators with varying lengths from 300 $\mu$m down to 50 $\mu$m and representative spectra. Fitting N=6 resonators for each condition, Supplementary Fig. \ref{figs6}d shows little change in the Q factor with varying waveguide length. Resonators with $\Delta$w = 50 and 30 nm maintain high Q factors exceeding 1000 even in resonators down to 50 $\mu$m. Each resonator could potentially be further scaled down with added dielectric mirrors patterned on the waveguide ends to reduce scattering losses. Thus, it is possible to envision individual free space coupled high Q resonators on the order of a few $\mu$m.

\section*{Supplementary Note 5: Refractive index sensing figure of merit}

Affinity based sensors that rely on spectral shifts induced by environmental refractive index shifts are often evaluated by a Figure of Merit (FOM):

\begin{center}
    \begin{equation*}
        FOM = \frac{(\Delta\lambda/\Delta n) * Q}{\lambda_0}
    \end{equation*}
\end{center}

where $\Delta\lambda/\Delta n$ s the resonant wavelength shift induced by a change in the background medium refractive index or bulk refractive index sensitivity, Q is the resonator quality factor, and $\lambda_0$ is the resonant wavelength. To determine the bulk refractive index sensitivity of our devices, we take a series of optical measurements in various saline solutions with differing concentrations of NaCl dissolved in deionized water. Increasing NaCl concentrations have been shown to increase the refractive index of water.\cite{Saunders2016-hb} In Supplementary Fig.\ref{figs7}, the change in resonant wavelength indicates our sensors have a sensitivity of $\Delta\lambda/\Delta n$ = 270 nm/RIU. The sensitivity of the resonators does not vary significantly with $\Delta$w, as the block asymmetry alters the Q factor, but the modal overlap with the surrounding medium does not change. Given our resonator Q factors of $\sim$2200 (Main text Fig. 2), we obtain a sensing FOM of around 400. This value is larger than previous demonstrations in plasmonic or dielectric metasurface based sensors.\cite{Conteduca2021-hg,Yang2014-ta,Wang2021-zr,Yesilkoy2019-wz,Triggs2017-su} We also note that modification of our metasurfaces, such as introducing subwavelength gaps or slots along the waveguide that further expose the resonant mode to target analytes, could dramatically improve the sensitivity and FOM of future iterations of the devices.

\section*{Supplementary Note 6: Modeling self-assembled monolayers}

To estimate the resonant wavelength shifts corresponding to successive molecular layers deposited on our sensors (Main text Fig. 3c), we model each surface step with a thin dielectric shell and numerically calculate the response with FDTD as described above. The dielectric shells extend from all exposed silicon faces of our metasurface nanoblocks as shown in Supplementary Fig. \ref{figs8}. The initial bare sensor before surface functionalization is calculated with a 4 nm silicon dioxide layer due to the thermal passivation step performed after nanofabrication of our sensors.\cite{krzeminski2007} The AUTES layer is assumed to bind as a uniform monolayer of thickness 1.8 nm and refractive indices ranging from n = 1.40-1.45, based on reported literature values. The refractive index is calculated for a range that corresponds to typical estimated optical properties of biomolecular layers. The MBS layer is estimated as a 0.7 nm thick layer based on the reported spacer arm length of the molecule and refractive index of n = 1.40-1.45. The thiolated ssDNA probe layers are calculated with optical properties n = 1.37-1.382, accounting for potential reduction in density due to dilution with PEG chains.\cite{schreck2015,gao2006,pinto2020} We estimate the structure of our single-stranded DNA probes using the open source software “mfold” to predict the secondary structure in a 1X PBS solution and the most stable conformation is shown in  Supplementary Fig. \ref{figs9}. The probe layer thickness is estimated by the wormlike chain model\cite{sim2012}:

\begin{center}
    \begin{equation*}
        R_g^2 = \left(\frac{l L_p}{3}\right) - L_p^2 + \left(\frac{2 L_p^3}{l}\right) - \left(\frac{2 L_p^4}{l^2}\right)(1-e^{-l/L_p})
    \end{equation*}
\end{center}

where the monomer spacing a = 0.6 nm, persistence length $L_p$ = 1 nm, and contour length l = Na (N = number of nucleotides) give a radius of gyration, Rg, and layer thickness of $\sim$ 4 nm.\cite{mantelli2011} The dsDNA probe refractive index is approximated as n = 1.4-1.423 as estimated from the densification and increased polarizability of duplexed DNA compared to single stranded DNA.\cite{elhadj2004} The thickness is similarly estimated by the wormlike chain model, but with a = 0.3 nm and Lp = 30 nm, which returns a similar radius of gyration of 4 nm. Based on the short fragment length of our probe and targets at 22-26 nt, we do not expect a significant monolayer thickness change upon DNA hybridization.

\section*{Supplementary Note 7: Fluorescence microscopy}

Fluorescence experiments were performed after DNA hybridization experiments with target nucleic acids tagged with ATTO590 dye on the 5’ end. Dried samples were placed in a Zeiss AxioImager system and imaged with a 20x objective. Fluorescence images were acquired with 1000 ms exposures on a Zeiss Axiocam 506 mono camera. Fluorescence intensity values were averaged over a 80 x 40 $\mu$m area and were normalized to the maximum intensity values from chips hybridized with complementary E gene 
targets as seen in Supplementary Fig. \ref{figs10}.

\section*{Supplementary Note 8: Langmuir adsorption model}

Concentration dependent resonant wavelength shift responses in main text Fig. 4 were fit to the Hill equation (formally equivalent to the Langmuir isotherm):

\begin{center}
    \begin{equation*}
        \theta = \frac{\theta_{max}*X^h}{K_d^h+X^h}
    \end{equation*}
\end{center}

where $\theta_{max}$ is the saturated maximum binding signal at high target concentrations, X is the target concentration, h is the Hill coefficient which describes the slope of the curve, and $K_d$ is the concentration value that corresponds to half-maximum binding signals.
Furthermore, the Langmuir adsorption model is also used to fit the time varying responses in main text Fig. 5. The wavelength shift response as a function of time is described as\cite{noll2018}:

\begin{center}
    \begin{equation*}
        \theta(t) = \theta_{eq}(1-e^{-kt}) 
    \end{equation*}
\end{center}

where $\theta_{eq}$ is the saturated equilibrium binding signal and k is an "observed" rate constant that accounts for both target hybridization and reversible dehybridization rates.

\clearpage

\begin{figure}[htp!]
\centering
\includegraphics[width=0.5\textwidth]{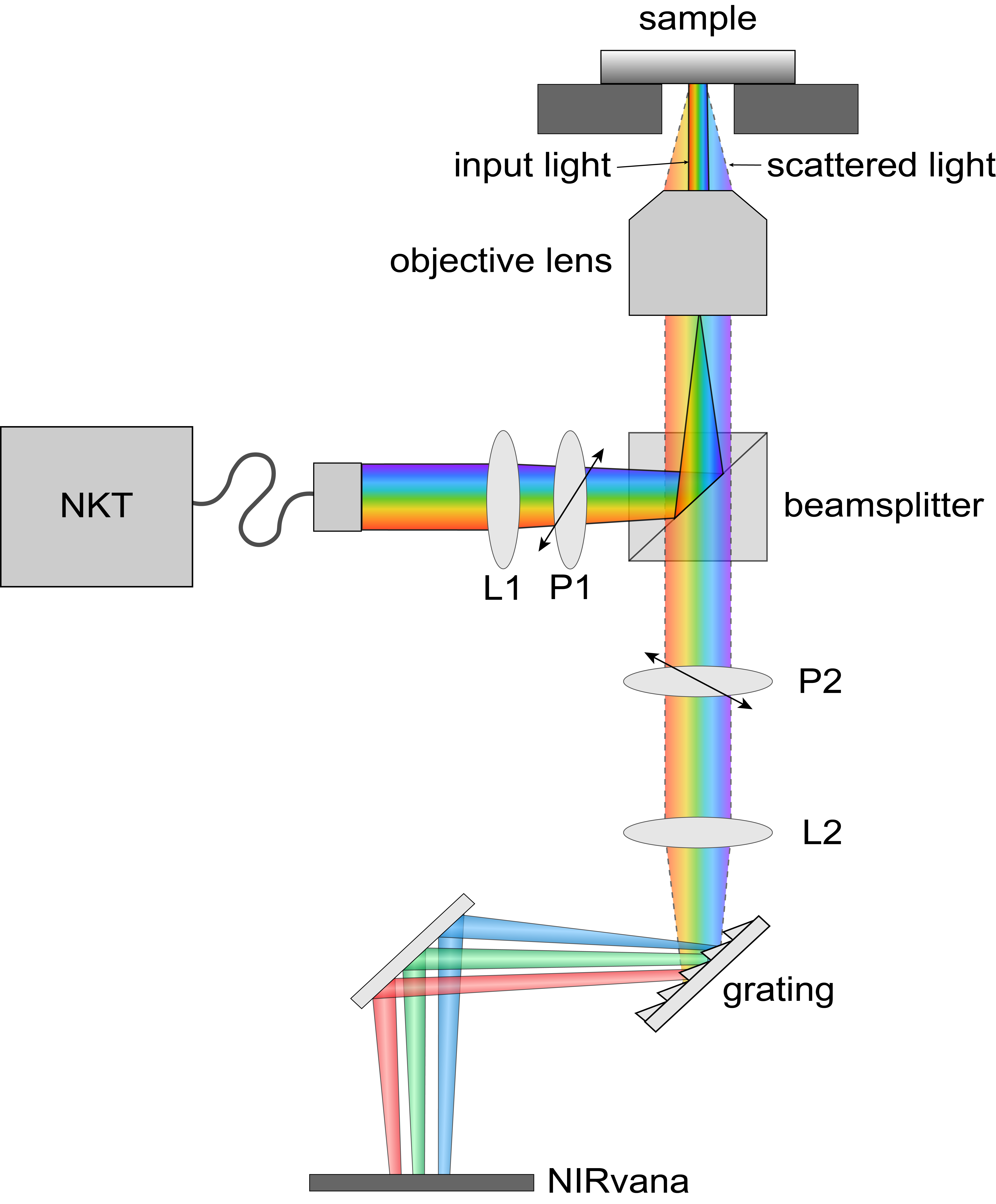}
\caption{\label{figs1} Schematic of near-infrared microscope set up utilized to collect spectra from metasurface samples.}
\end{figure}
\clearpage

\begin{figure}[htp!]
\centering
\includegraphics[width=1\textwidth]{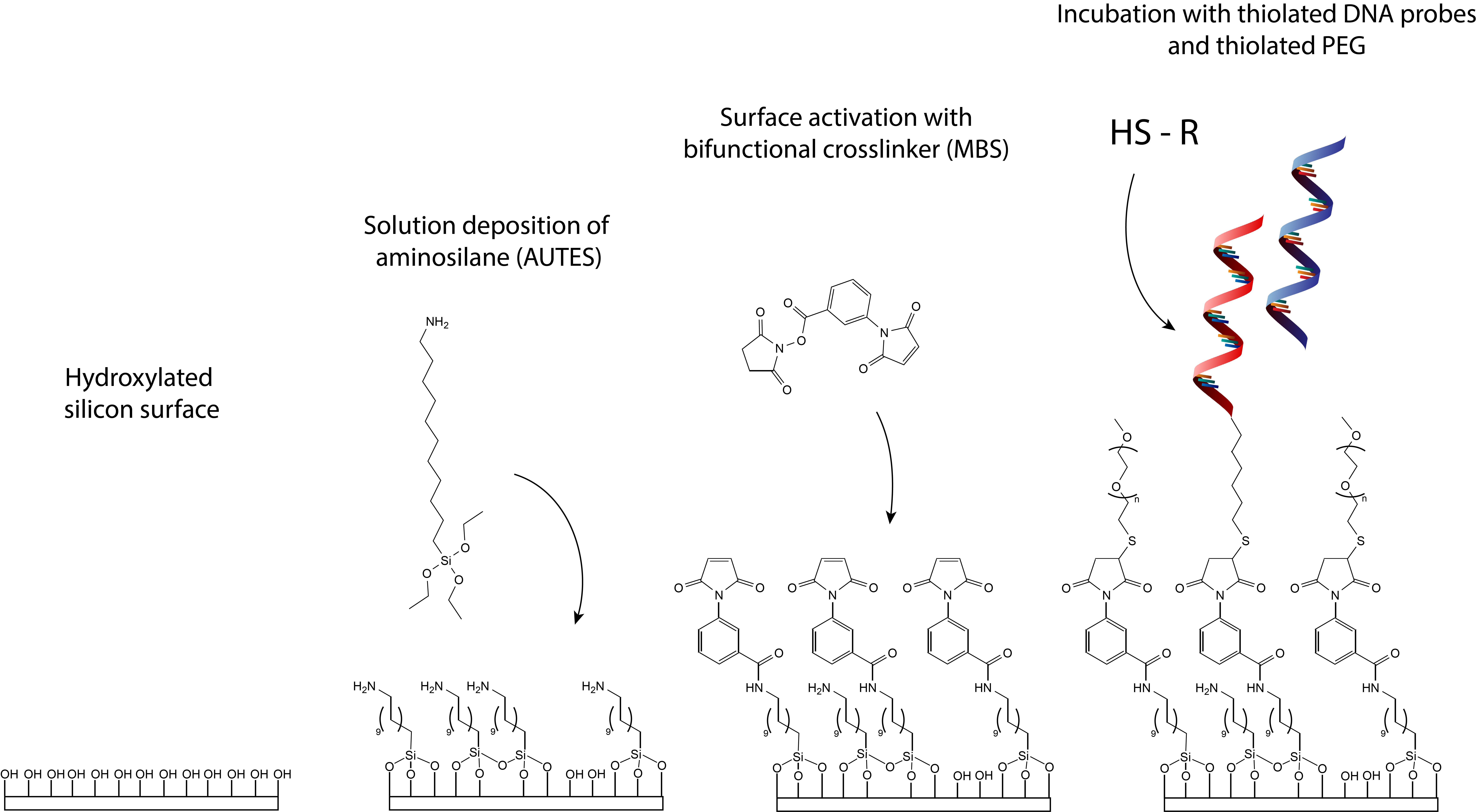}
\caption{\label{figs2} Multi-step surface functionalization of a DNA probe monolayer on silicon metasurfaces. }
\end{figure}
\clearpage

\begin{figure}[htp!]
\centering
\includegraphics[width=1\textwidth]{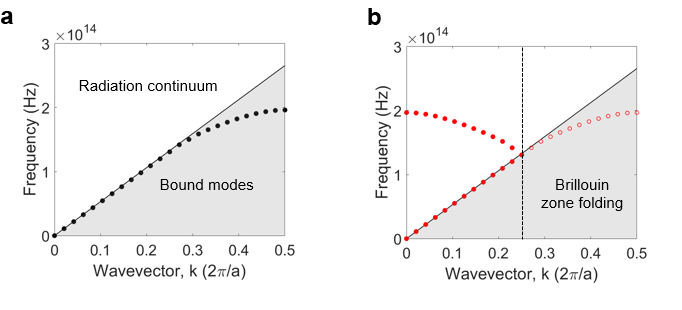}
\caption{\label{figs3} \textbf{a}, Simulated waveguide dispersion in an unperturbed chain of subwavelength silicon blocks. \textbf{b}, Brillouin zone folding introduced via symmetry breaking in biperiodic guided mode resonator.
 }
\end{figure}
\clearpage

\begin{figure}[htp!]
\centering
\includegraphics[width=1\textwidth]{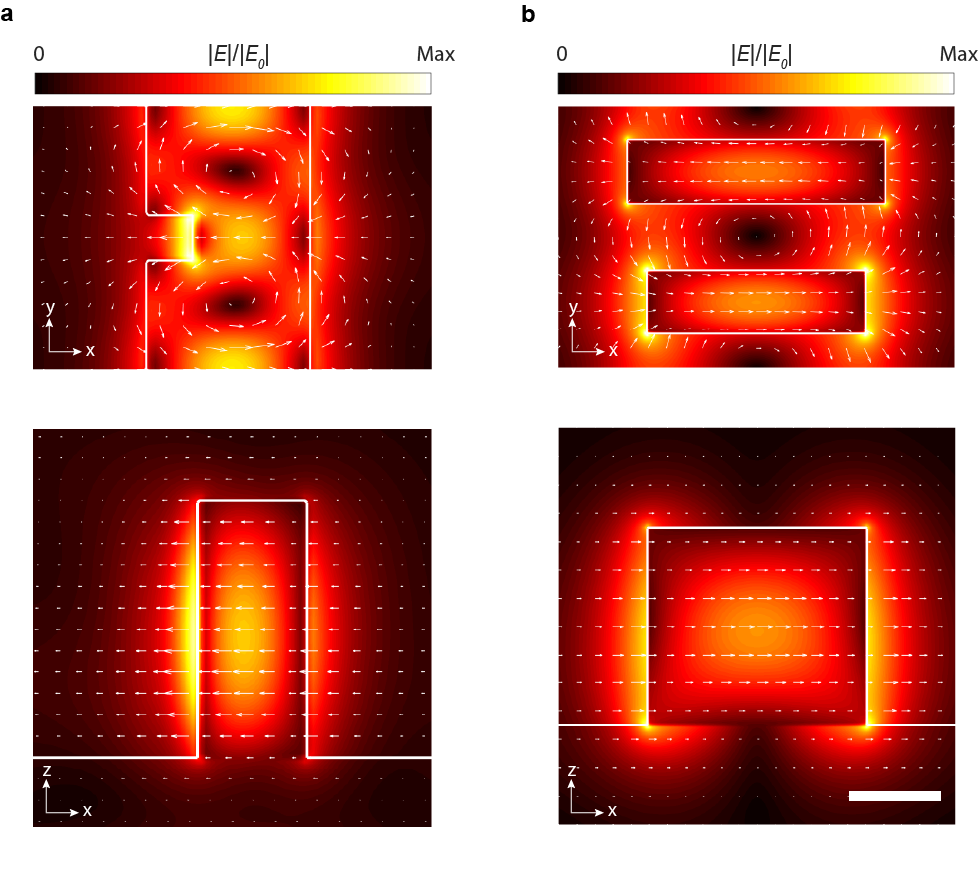}
\caption{\label{figs4} \textbf{a}, Electric field profile for notched silicon waveguide on a sapphire substrate. Upper panel shows an x-y cut through the center of the structure and the lower panel is an x-z cut through the center of the notch perturbation where fields are most strongly concentrated. \textbf{b}, Electric field profile for asymmetric chain of silicon blocks on a sapphire substrate. Upper panel represents the x-y cut through the center of the structure and the lower panel is an x-z cut through the center of the smaller block. Scale bar is 200 nm.
 }
\end{figure}
\clearpage

\begin{figure}[htp!]
\centering
\includegraphics[width=1\textwidth]{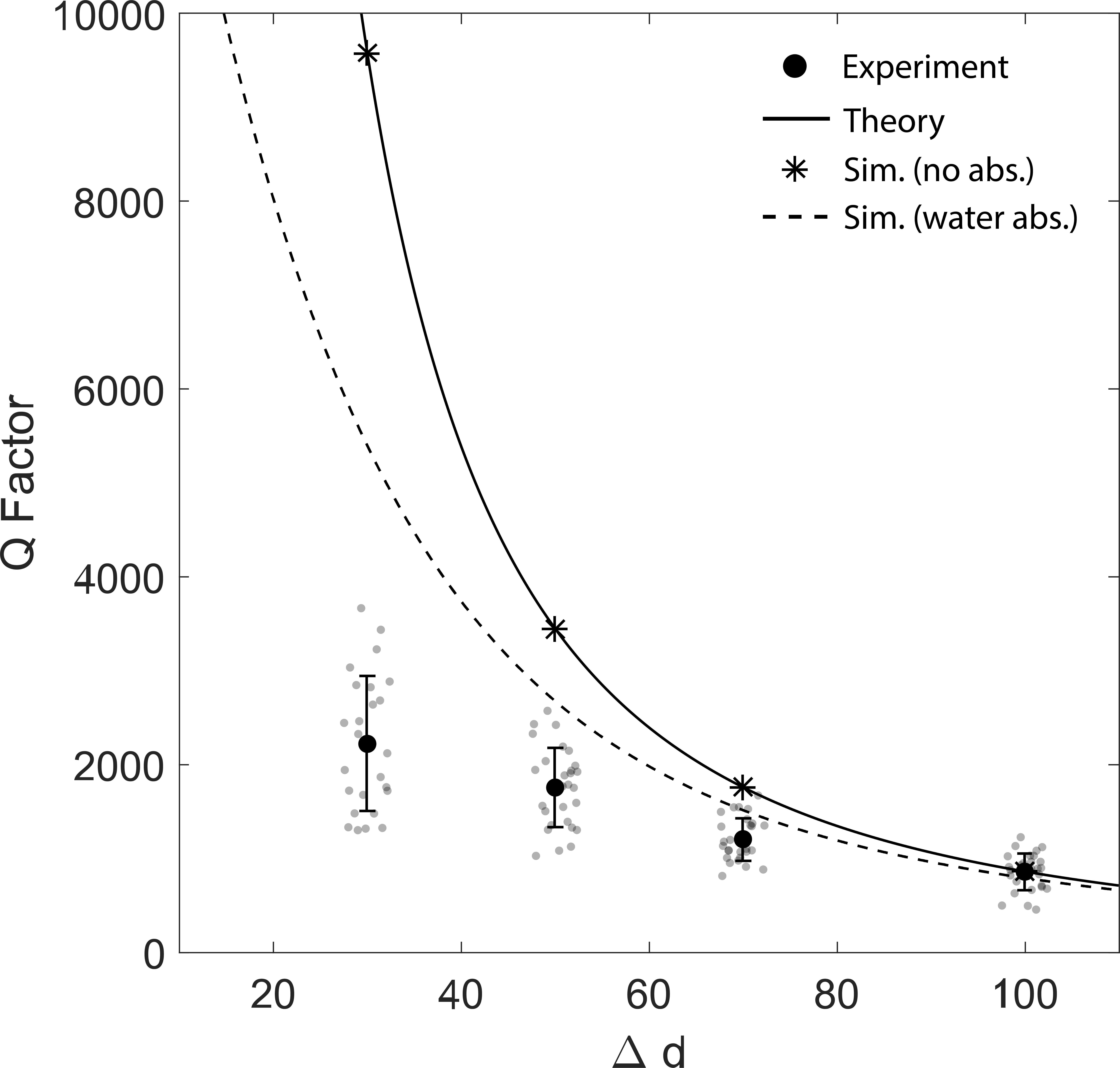}
\caption{\label{figs5} Resonator quality factor as a function of difference in neighboring silicon block length, $\Delta$d.
 }
\end{figure}
\clearpage

\begin{figure}[htp!]
\centering
\includegraphics[width=1\textwidth]{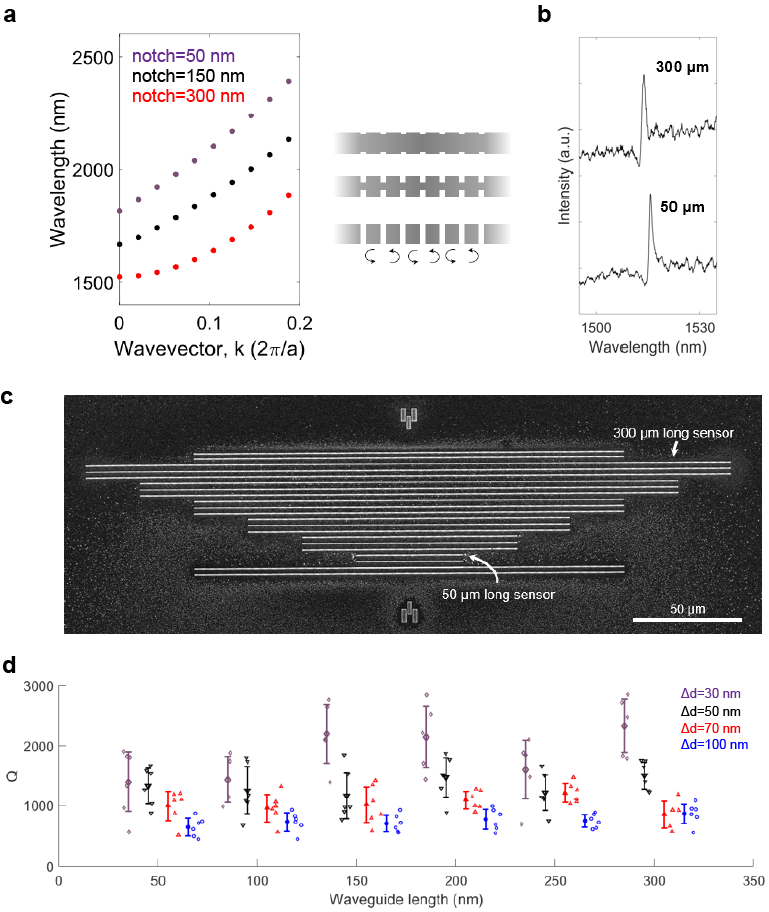}
\caption{\label{figs6} \textbf{a}, Simulated waveguide dispersion in waveguides with varying notch depth. \textbf{b},Representative optical spectra from waveguides with length of 300 $\mu$m and 50 $\mu$m. \textbf{c}, SEM image of resonators fabricated with different waveguide lengths. \textbf{d}, Quality factors as a function of waveguide length.
 }
\end{figure}
\clearpage

\begin{figure}[htp!]
\centering
\includegraphics[width=1\textwidth]{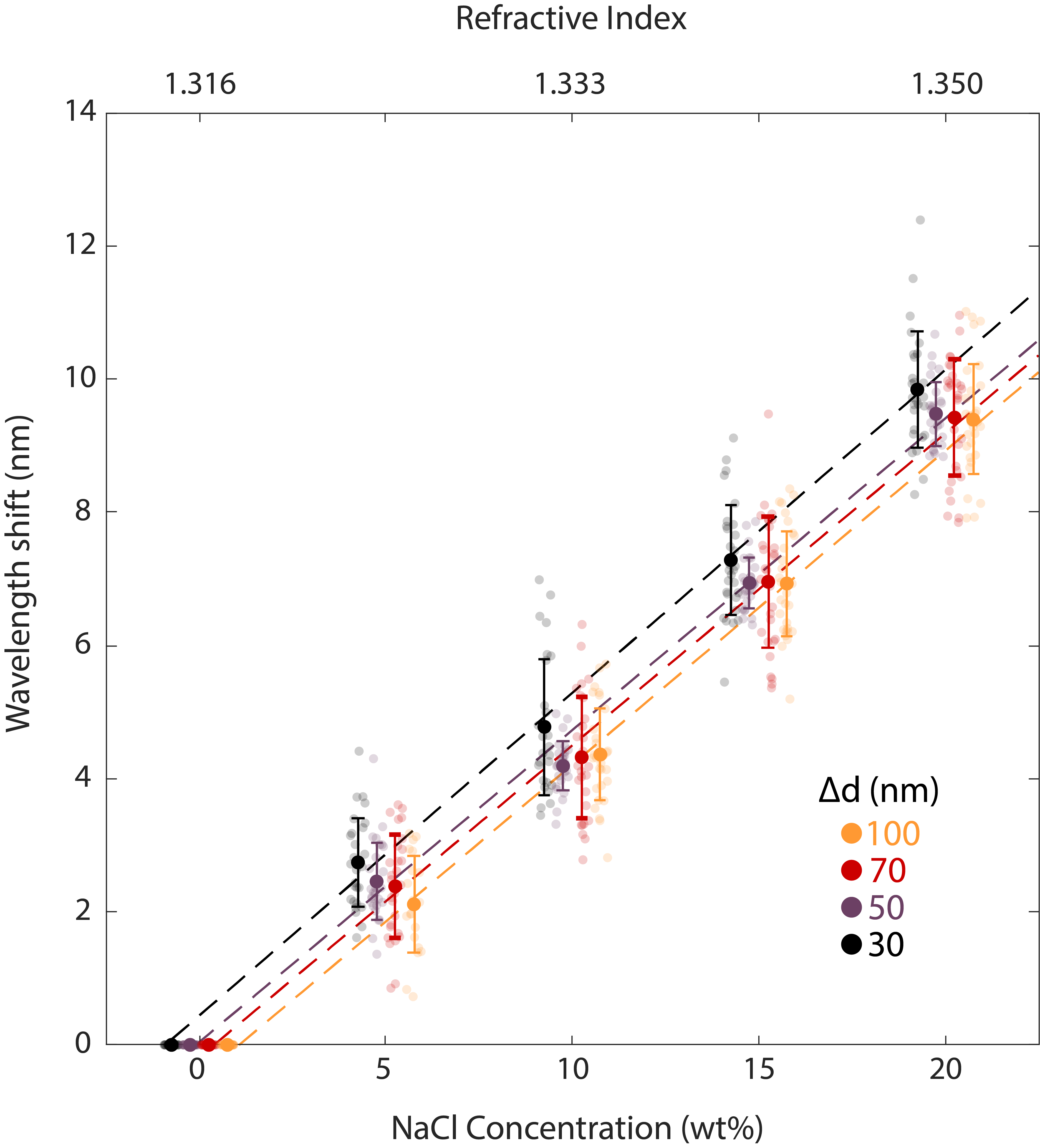}
\caption{\label{figs7} Resonant wavelength measurements as a function of background medium refractive index. Circles and error bars represent experimental measurements for N=25-30 resonators at each condition. Dashed lines represent linear fits to the data.
 }
\end{figure}
\clearpage

\begin{figure}[htp!]
\centering
\includegraphics[width=1\textwidth]{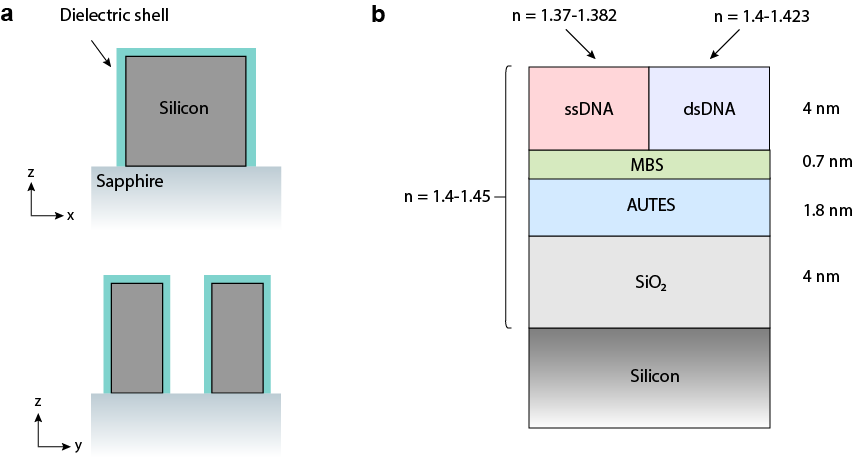}
\caption{\label{figs8}\textbf{a}, Schematic of calculated structure with dielectric shell around silicon nanostructures representing the molecular monolayers. \textbf{b}, Estimated layer thicknesses and refractive indices.
 }
\end{figure}
\clearpage

\begin{figure}[htp!]
\centering
\includegraphics[width=0.5\textwidth]{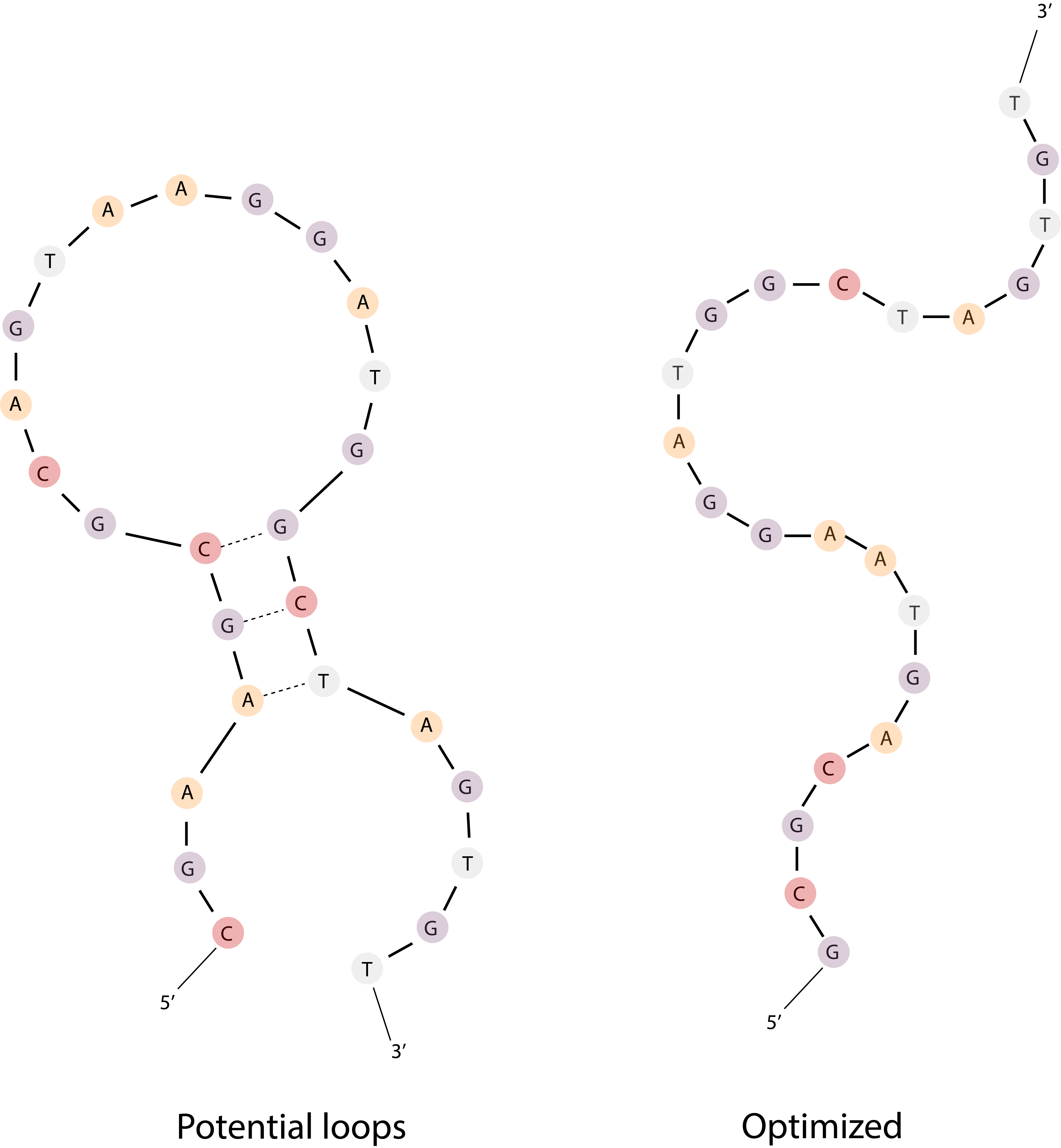}
\caption{\label{figs9} Likely secondary structure of DNA probes used in this work. Probes sequences were optimized to avoid formation of stable loop structures.
 }
\end{figure}
\clearpage

\begin{figure}[htp!]
\centering
\includegraphics[width=0.8\textwidth]{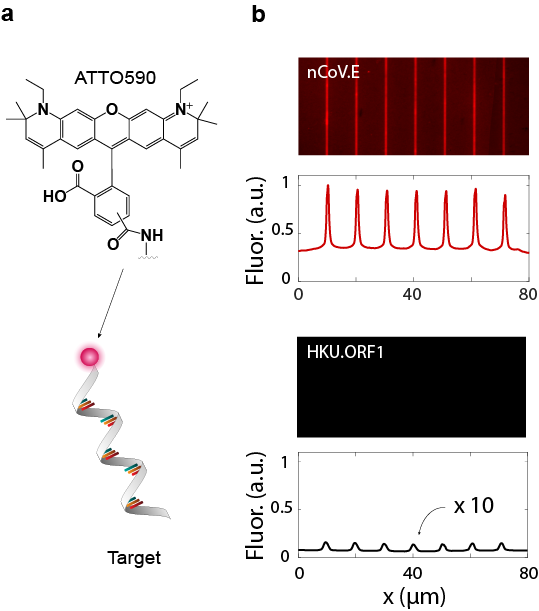}
\caption{\label{figs10} \textbf{a}, Schematic of fluorescently tagged target DNA sequences. \textbf{b}, Fluorescence images and integrated intensities for sensors exposed to complementary nCoV.E sequences (top) and non-complementary HKU.ORF1 sequences (bottom). Fluorescence imaging confirms the specificity of immobilized DNA probe molecules to complementary nucleic acid sequences. All metasurface sensors were functionalized with probes complementary only to the nCoV.E sequence.
 }
\end{figure}
\clearpage

\begin{center}
\begin{table}[htp!]
\begin{tabular}{ |c|c|c| } 
 \hline
 E gene Probe & 5' - G CGC AGT AAG GAT GGC TAG TGT - 3'\\
 E gene Target & 5' - ACA CTA GCC ATC CTT ACT GCG C - 3'\\
 ORF1b gene Target & 5' - CT AGT CAT GAT TGC ATC ACA ACT A - 3'\\
 \hline
\end{tabular}
\caption{\label{tab1} Probe and target sequences used in this work. }
\end{table}
\end{center}

\FloatBarrier
\pagebreak
\bibliographystyle{naturemag}
\bibliography{suppref}